# The Vulnerability of the Neural Networks Against Adversarial Examples in Deep Learning Algorithms

Rui Zhao
Faculty of Engineering and IT
University of Technology Sydney
Sydney, Australia
rui.zhao-2@student.uts.edu.au

*Abstract*—With the further development in the fields of computer vision, network security, natural language processing and so on so forth, deep learning technology gradually exposed certain security risks. The existing deep learning algorithms cannot effectively describe the essential characteristics of data, making the algorithm unable to give the correct result in the face of malicious input. Based on current security threats faced by deep learning, this paper introduces the problem of adversarial examples in deep learning, sorts out the existing attack and defense methods of black box and white box, and classifies them. It briefly describes the application of some adversarial examples in different scenarios in recent years, compares several defense technologies of adversarial examples, and finally summarizes the problems in this research field and prospects its future development. This paper introduces the common white box attack methods in detail, and further compares the similarities and differences between the attack of black and white boxes. Correspondingly, the author also introduces the defense methods, and analyzes the performance of these methods against the black and white box attack.

*Keywords*—Deep Learning, Security, Black box, White box, Adversarial Examples

## I. INTRODUCTION

Deep learning technology has been widely used in security sensitive tasks such as the face recognition system [1], autonomous driving [2], and security monitoring [3]. Attackers will analyze the vulnerability of the target model by certain means and design corresponding attack algorithm to maliciously tamper with the input samples, so as to reduce the performance of the target model. At present, the mainstream attack algorithm is mainly gradient computing. For example, the Fast Gradient Sign Method (FGSM) [4] constructs attack samples by calculating the gradient of the target model's loss function and looking for adversarial perturbation along the gradient direction, then adding it to the original normal input samples. I-FGSM and MI-FGSM are multi-step iterations of FGSM to obtain more aggressive attack samples. Before calculating the gradient of the loss function of the target model, R+FGSM adds Gaussian noise to the original input samples. The Carlini [5] attack algorithm uses gradient descent to optimize the objective function to construct attack samples. The design of these algorithms requires the attacker to have a full understanding of the target model structure or training data, that is, the white box attack. Obtaining the target model's internal information is, however, difficult in reality, which makes it a black box for attackers. Previous studies have shown that there is migration between different learning models, which means that attack samples generated by different attack algorithms can make multiple models misclassify at the same time. This feature lays a foundation for attackers to realize black box attack with target model's unknown internal information. The existing black box attack methods are mainly divided into transferability-based method [6-10], gradient estimation-based method [11-14], decision-based attack method [15] and sampling-based method [16]. This paper analyzes and discusses the security of black box attack and defense in deep learning algorithm.

Fig.1. This is an oppositional example created through using FGSM [4] on GoogleNet [17]. The interference generated by FGSM made GoogleNet recognize the picture from the panda to a gibbon, although it is pretersensual.

### A. Definition of black/white box attack

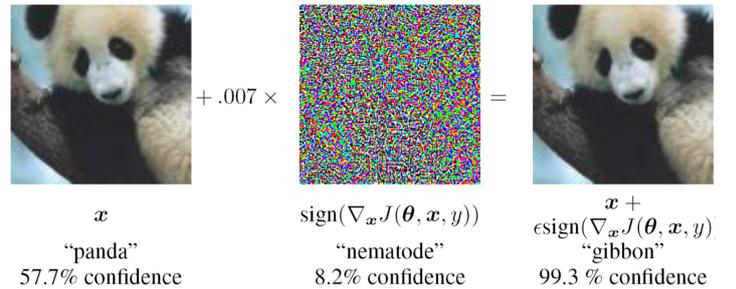

$x$
"panda"
57.7% confidence

$\text{sign}(\nabla_x J(\boldsymbol{\theta}, \boldsymbol{x}, y))$
"nematode"
8.2% confidence

$x + \epsilon\text{sign}(\nabla_x J(\boldsymbol{\theta}, \boldsymbol{x}, y))$
"gibbon"
99.3 % confidence

Based on the threat model, the existing adversarial attacks can be separated to various color attacks, such as white box attack, gray box attack and black box attack. The three models differ in the information the attacker knows. The white box attack model assumes that the attacker has full data of the target model, including its structure, parameters and so on. Therefore, attackers straight create adversarial examples on the target model as it may. The grey box supposes that the information the attacker knows is peculiar to the target model's structure and the query access right. In the black box, attackers can only rely on the returned results of query access to create adversarial examples. Guided by the threat models, researchers have developed certain attack algorithms to create adversarial examples, including limited-memory Broyden-Fletcher-Goldfarb-Shan-no (L-BFGS) and fast gradient sign method (FGSM) [17], Jacobian-based saliency map attack (JSMA) [7-8]. Although the attack algorithms were originally drafted under the white box threat model, it can be seen from the transitivity of the adversarial examples between the models that they are also applicable to the grey box and black box threat models.

## II. METHODS FOR GENERATING ADVERSARIAL EXAMPLES

### A. White box attack methods

#### 1) Box-constrained L-BFGS

Any few undetectable disturbances interfere to an image leads the idea the neural network could make misclassification, Szegedy [18] et al. primarily has been proved. Trying to solve the equation of minor disturbance which causes the neural network to make is not a proper solution by its high complexity, so the team changed the way to make problem simpler– finding the minimum loss function addition item, and making the neural network make misclassification. The solution performs a convex optimization process.

$$min_{x'} \quad c \parallel \eta + J_\theta(x', l')$$
$$s.t. \quad x' \in [0,1] \quad (1)$$

#### 2) Fast Gradient Sign Method (FGSM)

Finding the deep neural networks' robustness can build up adversarial training, Szegedy [17] has explored further on enhancing the ability to defend against attacks. GoodFellow [4] and others have developed a method that can effectively calculate against adversarial perturbation, called FGSM in the original context. An improved model of FGSM "one step target level" is proposed from Kurakin [19] et al. By substituting the class variables with the target type of smallest recognition probability in the adversarial perturbation, and subtracting the perturbation from the original figure, which would become the adversarial example and the target type can be output.

$$\eta = \epsilon sign(\nabla_x J_\theta(x, l)),$$

$$g_{t+1} = \mu g_t + \frac{\nabla_x J_\theta(x'_t, l)}{\parallel \nabla_x J_\theta(x'_t, l) \parallel}$$

$$x' = x - \epsilon sign(\nabla_x J(\theta, x, l'))$$

$$x_{tmp} = x + \alpha \cdot sign(\mathcal{N}(0^d, I^d)),$$

$$x' = x_{tmp} + (\epsilon - \alpha) \cdot sign\left(\nabla_{x_{tmp}} J(x_{tmp}, l)\right) \quad (2)$$

#### 3) Basic & Least-Likely-Class Iterative Methods

The one-step method demonstrates photo disturbance through adding the classifier's loss function in a huge step, so as to straightforward extend it to a variant of increasing the loss function into numerous tiny procedures, leading to Basic Iterative Methods (BIM). Variations of this approach share some in common with the previous one, which obtains the Least-Likely-Class Iterative Methods by substituting the class variables with the target type of smallest recognition probability in the adversarial perturbation. [19]

$$x_0 = x,$$

$$y_{LL} = argmin_y\{p(y|x)\},$$

$$x_{n+1} = Clip_x\{x_n - \epsilon sign(\nabla_x J(x_n, y_{LL})\} \quad (3)$$

#### 4) Jacobian-based Saliency Map Attack (JSMA)

The method generally has used in the literature of adversarial attack is to limit the value of the $l_\infty$ or $l_2$ canonical of the perturbation, therefore the perturbation in the adversarial example would not be noticed. To avoid affect the whole picture, JSMA just adjusts a few pixels. Papernot [7-8] et al. suggest the idea of evaluating the most sensitive input features from the Jacobian matrix by utilize JSMA method and find the marked pixels in the full photo that were useful to the realization of the attack target. By calculating the positive derivative (Jacobian matrix), this approach searches for marked points in order to recognize the input features that lead major changes in the output of the DNN. Compared with the inverse gradient FGSM calculation, the algorithm alters the original image pixel as a point each time, monitoring how its modifications influence the classification results. The calculation of the significance list is realized by using the gradient output of the network layer. After that, the most effective pixel is chosen in order to trick the network. The method of JSMA makes less modification on the original input and its calculation is comparatively easy, for JSMA applies forward propagation to the calculation of the protrusion.

$$J_F(x) = \frac{\partial F(x)}{\partial x} = \left[\frac{\partial F_j(x)}{\partial x_i}\right]_{i \times j} \quad (4)$$

#### 5) Carlini and Wagner Attacks (C&W)

Carlini and Wagner put forward three adversarial attack approaches, which limit the $l_\infty$, $l_2$ and $l_0$ norms to make the disturbance undetectable. Defensive distillation is impossible to defend against C&W's attacks by many experiments. The adversarial perturbation created by this algorithm can shift the network from an unsecured one to a secured one, accordingly recognizing black box attacks.

Most existing adversarial detection defenses are effective against C&W's Attack. [20-21]

$$\min_\eta \parallel \eta \parallel_p + c \cdot g(x + \eta)$$

$$s.t. \quad x + \eta \in [0,1]^n$$

$$g(x') = max\left(\max_{i \neq l'}(Z(x')_i) - Z(x')_t, -k\right)$$

$$\min_w \parallel \frac{1}{2}(tanh(w) + 1) \parallel_2 + c \cdot g(\frac{1}{2}(tanh(w) + 1))$$

$$\min c \cdot g(x + n) + \sum_i[(\eta_i - \tau)^+] \quad (5)$$

#### 6) DeepFool

Moosavi-Dezfooli [22] et al. used iterative calculation to generate adversarial perturbation with the minimum norms. The image which was originally placed within the classification boundary is pushed out of the territory step by step before misclassification occurs. They produce tinier disturbances with similar deception rates.

$$\underset{\eta_i}{argmin} \ \|\eta_i\|_2$$

$$s.t. \ f(x_i) + \nabla f(x_i)^T \eta_i = 0 \quad (6)$$

*7) Universal Adversarial Perturbations*

Compare to Universal Adversarial Perturbations [23] are able to generate perturbations that can attack any image and perturbations are almost invisible at the same time, the adversarial perturbation can only be produced as one single image through FGSM, ILCM, and DeepFool methods. As shown in the figure below, this general perturbation can make a classifier misclassify all images.

$$\|\eta\|_p \leq \epsilon,$$

$$\mathcal{P}(x' \neq f(x)) \geq 1 - \delta. \quad (7)$$

*8) Adversarial Transformation Networks (ATNs)*

Baluja and Fischer [24] trained multiple feed forward neural networks to produce adversarial samples are able to attack one or more networks. The algorithm generates adversarial examples through the minimization of a joint loss function composed of two portions. The first portion keeps the adversarial example similar to the original image; the second portion misclassify the adversarial example.

*9) CPPN EA Fool*

The Compositional Pattern-Producing Network-encoded EA (CPPN EA) method uses an evolutionary algorithm (EA) to generate adversarial examples, which generated by this method will be misclassified by DNNs with 99% confidence. An updated attack- CPPN EA- was discoverd by Nguyen et al. that is impossible for human to recognize because adversarial examples use deep neural networks with high confidence (99%) to classify [25].

## III. METHODS OF BLACK BOX ATTACK

### A. One Pixel Attack Gradients

An extreme form of adversarial attack. Changing a single pixel of the image could achieve an attack. Su et al. iteratively altered every pixel to produce a sub-image making comparison with parent image as a differential evolution algorithm. Keep the sub-images that attack best according to the selection criteria to realize the attack against the enemy. Information of network parameters or gradients are not required for the adversarial.

Su et al. avoided the problem of measurement of perceptiveness by modifying a single pixel to generate adversarial examples. [26]

$$\underset{x'}{min} \ J(f(x'), l')$$

$$s.t. \ \|\eta\|_0 \leq \epsilon_0 \quad (8)$$

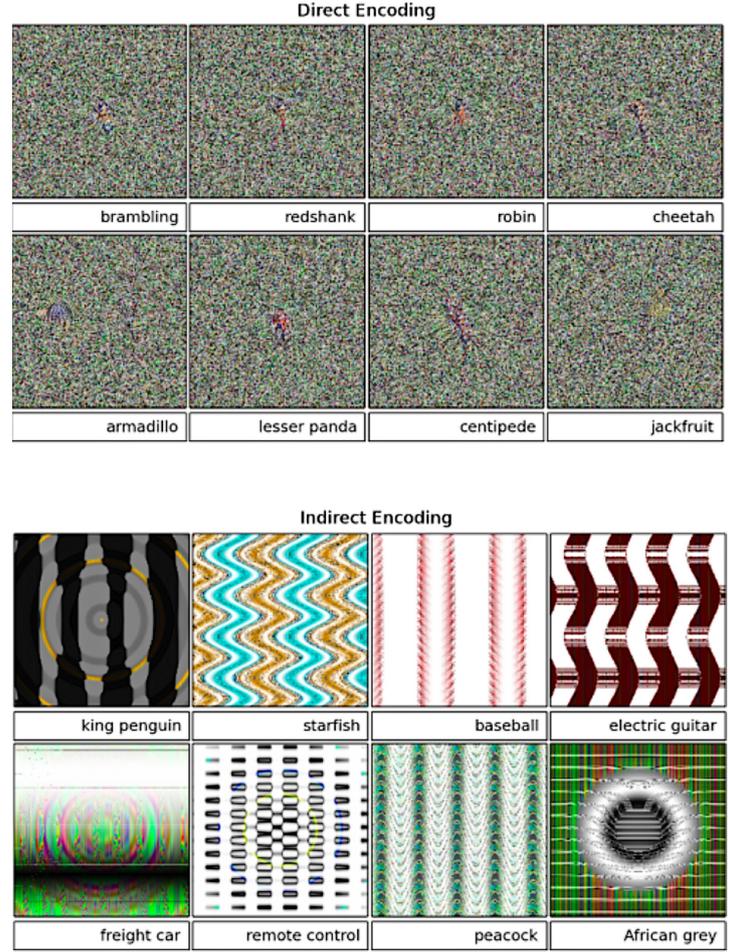

Fig.2. Examples that cannot be recognized by humans, but are classified by deep neural networks into a class with high certainty (≥99.6%).

### B. UPSET and ANGRI

UPSET and ANGRI are two black box attack algorithms proposed by Sarkar [27] et al. UPSET is able to produce adversarial perturbation for an exact target classification. By adding interferon, the image can be grouped into the target class. In contrast to the "image-agnostic" perturbation of UPSET, ANGRI produces "image-specific" perturbation, both of which have high trick rates on MNIST and CIFAR data sets.

### C. Natural GAN

Zhao [28] et al. employed the GAN method to create adversarial examples for the purpose of making the results more natural. This method generates adversarial noise by minimizing the internal distance such as "Feature Adversary". It is not necessary to use the gradient of the neural network in this method, which is regarded as a black box attack.

$$\underset{z}{min} \ \|z - \mathcal{I}(x)\|$$

$$s.t. \quad f(\mathcal{G}(z)) \neq f(x). \tag{9}$$

*D. Houdini*

Houdini [29] is such a method as can trick gradient-based machine learning algorithms. It generates countermeasures against examples that are particular for the loss function task. In other words, it produces adversarial perturbation by using the gradient information of the network's differentiable loss function. Furthermore, this algorithm is not only able to classify photo networks, but can also be used to trick speech recognition networks.

*E. Zeroth Order Optimization (ZOO)*

The ZOO method does not use gradient training to directly attack the model to generate adversarial examples. It is a black box attack. Because it is a black box attack, the gradient does not need to be calculated, but needs to be queried and evaluated. Based on this, Chen [30] et al. proposed the ZOO-ADAM algorithm, which is to stochastic select a variable to update the adversarial example. Trials show that the ZOO method achieves the same performance as the C&W attack method.

$$g(x') = \max \begin{pmatrix} \max \\ i \neq l' \end{pmatrix} (log[f(x)]_i) - log[f(x)]_{l'}, -\kappa),$$

$$\frac{\partial f(x)}{\partial x_i} \approx \frac{f(x+he_i)-f(x-he_i)}{2h},$$

$$\frac{\partial^2 f(x)}{\partial x_i^2} \approx \frac{f(x+he_i)-2f(x)+f(x-he_i)}{h^2} \tag{10}$$

*F. Comparative analysis of the black box and white box*

In actual deployment and application, we are usually unable to obtain the model's structure, parameters and other information, and can only manipulate the input and output of the model. Therefore, in this scenario, black box attacks are more common and practical. According to the different strategies used in the attack, the existing black box attack methods are mainly divided into transferability-based methods [6-10], gradient estimation-based methods [11-14], decision-based attack methods [15] and sampling-based methods [16].

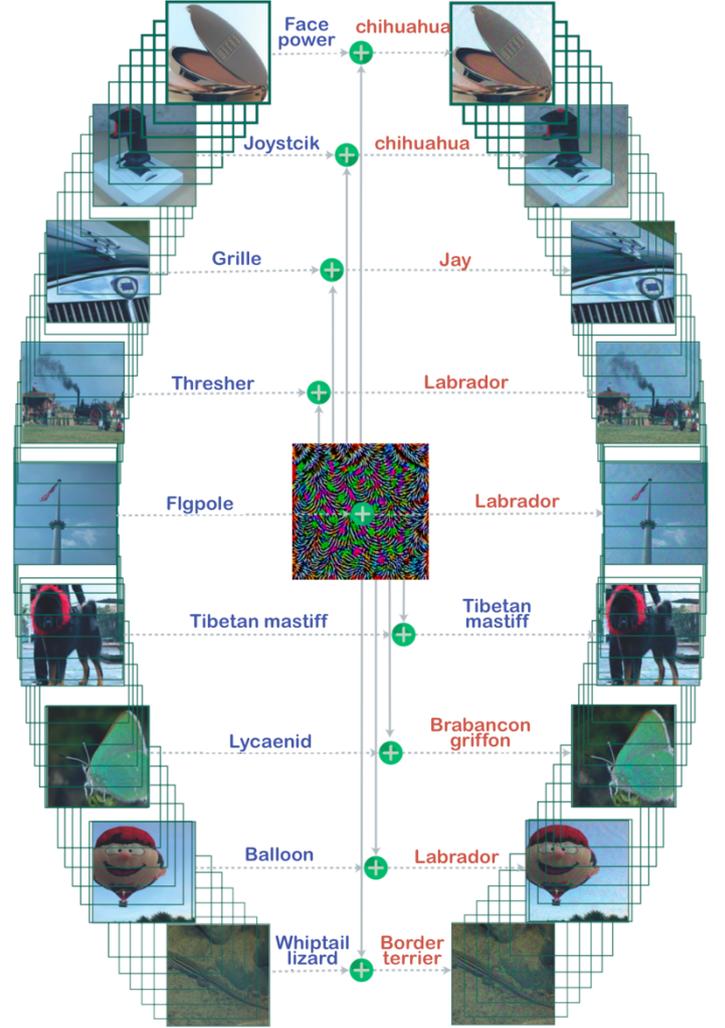

Fig.3. A universal adversarial example tricks the neural network on images. Image on the left: Natural images of the original markups. Central image: Universal perturbation. Image on the right: Confused images with wrong labels.

TABLE I. CLASSIFIES 9 WHITE BOX ATTACK PRACTICES AND 6 BLACK BOX ATTACK PRACTICES FROM FIVE ASPECTS OF TARGETED/NON TARGETED, SPECIFIC/UNIVERSAL, PERTURBATION, LEARNING, AND STRENGTH

| Method | Black/White | Targeted/Non-targeted | Specific/Universal | Perturbation | Learning | Strength |
|---|---|---|---|---|---|---|
| L-BFGS [18] | White Box | Targeted | Image specific | $l_\infty$ | One shot | *** |
| FGSM [5] | White Box | Targeted | Image specific | $l_\infty$ | One shot | *** |
| BIM&ILCM [19] | White Box | Non-Targeted | Image specific | $l_\infty$ | Iterative | **** |
| JSMA [8-9] | White Box | Targeted | Image specific | $l_0$ | Iterative | *** |
| C&W [20-21] | White Box | Targeted | Image specific | $l_\infty, l_2$ 和 $l_0$ | Iterative | ***** |
| Deep Fool [22] | White Box | Non-Targeted | Image specific | $l_\infty, l_2$ | Iterative | **** |
| Uni.perturbations [23] | White Box | Non-Targeted | Universal | $l_\infty, l_2$ | Iterative | ***** |
| ATNs [24] | White Box | Targeted | Image specific | $l_\infty$ | Iterative | **** |
| CPPN EA [25] | White Box | Targeted | Image specific | N/A | Iterative | *** |
| One pixel [26] | Black Box | Non-Targeted | Image specific | $l_0$ | Iterative | ** |
| UPSET [27] | Black Box | Targeted | Universal | $l_\infty$ | Iterative | **** |
| ANGRI [27] | Black Box | Targeted | Image specific | $l_\infty$ | Iterative | **** |
| Natural GAN [28] | Black Box | Non-Targeted | Universal | $l_2$ | Iterative | *** |
| Houdini [29] | Black Box | Targeted | Image specific | $l_\infty, l_2$ | Iterative | **** |
| ZOO [30] | Black Box | Non-Targeted | Image specific | $l_2$ | Iterative | ** |

*1) Transferability-based methods*

Research shows that the adversarial examples have transferability [6], that is, the adversarial examples generated for the target model may also make other models with different structures and training sets go wrong. Therefore, in the black box scenario, the attacker can train his own model on a data set that is the same as the black box target model or has a similar distribution, and the adversarial example generated by his self-trained model uses its transferability to deceive the black box target model. When the training data are not available for the attacker, he/she is able to use the target model to label the data synthesized by himself based on the idea of model distillation, and use the synthetic data to train the substitute model to get close to the target black box model. Then the attacker uses the white box attack method to produce adversarial examples against the substitute model, with which a black-box transferable attack on the target model can be performed [8]. However, though this method has been proven to be suitable for data sets with low intra-class differences, such as MNIST, there is no research to prove that it can be extended to more complex data sets such as CIFAR or ImageNet.

*Black box attack process*

Assume that $X \in R^{i*D}$ means the matrix composed by I samples with D feature spaces, and $Y=\{y^1, y^2, \cdots, y^c\}$ means the set of c different tags. Given a synthetic data set $Data_{\text{synthetic}}=\{(x_i, y_i)|1 \leq i \leq I\}$, $x_i \in X$ means the i training sample, and $y_i \in Y$ means the feedback tag of the i sample target model. The process of black box attack begins with learning a substitute model F(x)F(x) from $Data_{\text{synthetic}}$, and select a certain attack algorithm to generate an attack sample $x^{adv}$ for that model. Then, the sample is transferred to the target model O(x), resulting in the misclassification of the target model, that is O(x)≠O($x^{adv}$).

*2) Gradient estimation-based methods*

Chen et al. [11] proposed that adversarial examples could be generated by directly estimate the gradient of the target deep learning model with ZOO, the finite difference algorithm based on zero-order optimization. The experimental results presents that the ZOO attack algorithm is significantly superior than the black box attack algorithm on the basis of alternative models and is comparable to the effect of the white box algorithm C&W attack. However, this method requires more queries and depends on the predicted value of the model, such as category probability or confidence, so it cannot be applied to the situation of limited queries or when model only has a category label. For the limited number of doubts, Bhagoji et al. [12] applied Random Feature Grouping and principal component analysis (PCA) algorithms to make less number of queries required for generating adversarial examples. Ilyas et al. [13] combined Gradient Priors and Bandit Optimization algorithms to overcome this limitation. Tu et al. [14] proposed the AutoZOOM framework, which mainly includes two modules:

a) The adaptive stochastic gradient estimation strategy for balancing the number of model queries and distortion;

b) Autoencoder or bilinear adjustment operation of offline training with unlabeled data used to improve attack efficiency. When this framework is applied to the ZOO attack algorithm, it remarkable reduces the need of queries while maintaining the attack effect.

*3) Decision-based attack methods*

In practical machine learning-related applications, attackers can seldom obtain the predicted value of the model. For the situation where the target model only has a category label, Brendel et al. [15] proposed the boundary attack algorithm. Its main idea is to make the initialized image or noise gradually approach to the original sample until the decision boundary is found, and the closest adversarial example is found on this boundary. Compared with transferability-based attacks, they require less information and have simple implementation and greater practicability, but they need extensive of queries. In the presence of defense methods such as gradient masking, internal randomness or adversarial training, this decision-based attack is harder to defend than other types of black box attacks.

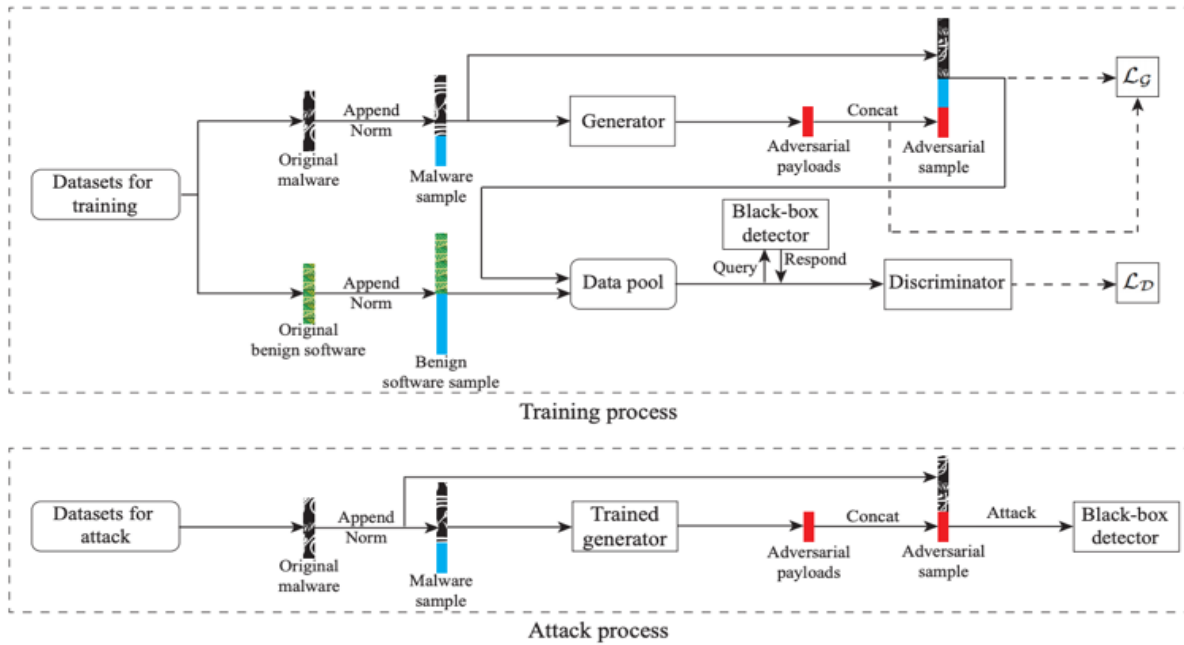

Fig.4. Flow chart of black-box attack

## IV. COMMON DEFEND METHODS

The current defense methods against adversarial examples mainly include entire defense and detection defense. The aim of the complete defense method is to enable the DNN classifier to recognize the adversarial example as the correct label after defense, while the detection defense method only demands to identify whether the input sample is the adversarial example, and not the actual label of the example itself. The complete defense method can be further divided into four types including adversarial training, gradient masking, input transformation and region-based classification. The following table summarizes current adversarial example defense methods.

### A. Complete defense

The adversarial training defense method begins with the training data set, and constantly provides the training process with adversarial examples so as to construct a more robust model to approach the goals of defending against adversarial examples. The various types of adversarial examples added to the training process allow researchers to propose many different defense methods based on adversarial training. For example, the ordinary adversarial training [31], PGD adversarial training [32] and ensemble adversarial training [33] defense methods are adding LLC, R+FGSM, and PGD, three adversarial examples generated by different attack algorithms, respectively during the training process. Generally speaking, most adversarial training defense methods increase adversarial examples in adversarial training, which are generated by attacking their own models. However, the gathering adversarial training defense method splits the training process from the generation process, and the adversarial examples added in the training process are generated by attacking other models, which advances the diversity of the adversarial examples added, thereby improving the ability to resist other attack methods.

### B. Gradient masking

Most attack algorithms utilize the model's gradient information to produce adversarial examples, and the gradient information is hidden by the gradient masking defense during model training. As a result, it is challenging for attack algorithms to use the gradient solution method to conduct its attack. The deep compression network method [34] introduces the smoothing penalty of the compressed autoencoder in the training process, so that the input can be more sensitive to the output and achieve the goal to hide the gradient information. Distillation is used in the distillation defense method [35] to train two cascaded DNN models to raise the robustness of model prediction. Input gradient regularization [36] penalizes the degree of change of the output to the input on the training's objective function, which can limit small adversarial perturbation to a certain extent in order not to drastically change the prediction results of the final model.

### C. Input transformation

The input transformation method attempts to reduce the perturbation that may exist in the sample to be predicted through various transformation methods, after which it directly inputs the converted sample into the original prediction model. The advantage of input transformation is that there is no need to change the data set for retraining, nor for the original model's structure. The ensemble input transformation method [37] integrates the five most commonly used image preprocessing and transformation methods, and directly performs five image transformations on the predicted samples concurrent, which improves the accuracy of the model's prediction of adversarial examples. Similarly, the input random transformation method [38] adds two additional layers of random transformation processes, including random resizing and filling, before the samples to be predicted are input into the original model, and then use the original model to predict. The thermometer encoding method [39] discretizes continuous input samples with thermometer encoding, so it uses the encoded samples in both training and prediction phases. Pixel Defense [40] uses the Pixel CNN to generate model to transform the adversarial examples into the regular sample space, after which the transformed sample is then fed into the original model to predict.

Region-based classification-The region-based classification randomly selects several samples from the region of the sample to be predicted and uses the original model to predict all samples. Then, it uses the majority voting method to select the most predicted one as the final label of the sample to be predicted. [41]

### D. Detection defense

Although the above-mentioned complete defense methods have received high expectation, many trials display that their actual influence is not good enough. Accordingly, considering the difficulty of those methods, analysts have put forward a collection of detection methods of adversarial examples. The detection defense method is only for the evaluation of whether or not the sample is an adversarial example. It does not have to identify its true label. The detection method based on local intrinsic dimension (LID) [42] uses the property that the LID value of the adversarial example is much larger than normal samples to identify the different samples. In the Feature Squeezing method [43], two external models are added into the DNN classifier in order to reduce each pixel's color bit depth and to make spatial smoothing of the pixel value. If the difference between the test sample and the sample processed by the external model after being predicted by the classifier is large, the test sample will be considered as an adversarial example. In fact, MagNet [44] includes two methods – complete defense and detection defense. MagNet first uses Detector to detect adversarial examples with large perturbation and discards them directly. Then, for those with small perturbation, Reformer is used to convert them into normal samples, and finally they are identified by the original model.

TABLE II. CLASSIFICATION OF ADVERSARIAL EXAMPLE DEFENSE METHODS

| | | |
|---|---|---|
| **Complete defense** | **Adversarial training** | General adversarial training [31], PGD adversarial training [32], ensemble adversarial training [33] |
| | **Gradient masking** | Deep compression network [34], Defensive Distillation [35], input gradient regularization [36] |
| | **Input transformation** | Ensemble input transformation [37], input random transformation [38], thermometer decoder [39], PixelDefence [40] |
| | **Classification** | region-based classification [41] |
| **Detection defense** | | Detection based on local intrinsic dimension [42], Feature Squeezing [43], MagNet [44] |

## V. ANALYSIS OF DEFENSE METHODS

Although current adversarial example defense methods have achieved certain effects, there still remain limitation and challenges.

### A. Adversarial training-based defense

On the one hand, the adversarial training defense method requires not only a lot of regular samples, but also extensive adversarial examples, with huge growth of the training time and required resources, making it difficult to use the defense on large-scale data sets in practice. On the other hand, since only limited adversarial examples generated by known attacks can be added to the training process, adversarial training defense is usually only effective for adversarial examples of the same type as the training, and does not has the ability to generalize different samples of other attacks.

### B. Gradient masking defense

Researchers believe that this defense method is easy to bypass, and even consider it a "failed" defense. A simple way to bypass this defense is that the attacker trains an alternative model similar to the model after defense and constructs an adversarial example by using the gradient of the alternative model to achieve the purpose of bypassing the model. Similarly, the gradient masking method requires changing the model structure and retraining the classifier.

### C. Input transformation-based defense

Generally, no network structure or training data sets are changed, but the prediction samples need to be converted. Theoretically, the input transformation method has a certain defensive effect on any kinds of attack samples, but experiments show that this defense method has a higher false alarm rate and missed alarm rate in the prediction of adversarial examples.

### D. Region-based classification

It mainly uses the predicted values of other samples in the region of the sample to be predicted. This method has two important parameters, namely "r", the radius of the region, and "N", the number of samples. However, the larger the "r", the lower the accuracy of the classifier in predicting normal samples. And the smaller the "r", the worse the defense effect of the method against adversarial examples. Similarly, with a larger "N", the efficiency of the method will increase exponentially. And the smaller the "N", the worse the defense effect will be.

### E. Detection defense

On the one hand, this kind of method will reduce the classification accuracy of normal samples. On the other hand, this method does not indicate the solution to the condition when the detection result is an adversarial example.

In general, current defense methods face various challenges. On the one hand, each defense method can only resist a limited number of adversarial examples, and can easily be bypassed by the examples that continuously evolve and mutate. On the other hand, the existing defense methods are passive defenses that can only defend against a certain type of attack and cannot solve unknown risks.

## VI. PROBLEMS AND PROSPECT OF CURRENT DEFENSE METHODS

With further promotion in diverse fields, deep learning technology has steadily become an important engine driving the accelerated development of economy and society from digitization and networking to intelligentization. Faced with the potential security threats brought by adversarial examples, this article first introduces a series of related concepts and properties of adversarial examples. Then it summarizes the explanations of the existence of adversarial examples, and summarizes existing problems. Moreover, it lists the classic black and white box attack methods. Finally, a series of adversarial example defense methods are sorted out.

### A. Major problems

the attach and defense of adversarial examples develop together and complement each other. Although the current research on adversarial examples defense methods has achieved certain results, there are still many challenges. The main problems are from three aspects.

*a)* The defense of adversarial attacks has a problem of dependence on the target model's parameters. The white-box defense strategy used by the model is to change the gradient transmission process of the target model, while the black-box attack uses alternative models to construct adversarial examples, and its own transferability properties give it good generalization property in black box attacks, which makes the white box defense strategy used by the model invalid.

*b)* Almost all defense methods can only be effective against limited adversarial attacks and cannot solve the risks from unknown attacks. They are easily bypassed by evolving adversarial examples.

*c)* Most defenses aim adversarial examples at computer vision tasks. As adversarial examples in other fields develop, there is an urgent need to study the problems in these fields. For example, in the field of cyberspace security, the biggest problem with some deep learning cyberspace security applications is poor robustness and is vulnerable to adversarial attacks.

### B. Prospect

Adversarial examples are a great threat in deep learning practice. Existing deep learning technologies generally face security challenges brought by adversarial examples. There are still problems that need to be solved to further deepen the research on the mechanism of adversarial examples, to expand more research on adversarial examples applied in more fields, and to build safe and credible deep learning models. Minimizing risks and ensuring the safe, reliable, and controllable development of artificial intelligence have great scientific significance and application value. Upcoming 5G network will push deep learning technology development forward by its massive data. Beyond that, it will bring up more security issues relate to adversarial cases that attract public attention especially academic scholars to solve. Combining with the problems that need to be solved in current adversarial example research field, this paper collects and summarizes the following research orientations in the adversarial example field.

*a)* Enhance model interpretability. The existing inexplicability of deep learning brings more business risks, and improving the interpretability of deep learning systems boosts better analysis of their logical vulnerabilities. Therefore, introducing useful mathematical tools to analyze deep learning models and constructing an integrated theoretical model for the cause of adversarial examples are the focus of this research field.

*b)* Improve the robustness evaluation system of deep learning. At present, the attack and defense of adversarial samples lack a unified and complete evaluation standard. The establishment of a united test data set will be a useful supplement to the research of adversarial examples. Therefore, building a universal and sound deep learning evaluation defense system around the integrity and usability of the model is an urgent problem to be solved in existing research on adversarial examples.

*c)* Introduce cryptographic methods. Data security and privacy protection are important components of deep learning systems. Correspondingly, the cryptography technology can provide powerful supplements and guarantees for deep learning systems. Therefore, while considering the efficiency, using cryptographic techniques - for instance, differential privacy protection and homomorphic encryption – so as to protect user data privacy and ensure the integrity and availability of data is a direction worth exploring in research of adversarial examples.